\documentclass{sample-algotel}
\usepackage[latin1]{inputenc}
\usepackage[francais]{babel}
\usepackage{float}
\usepackage{amsmath}
\usepackage{graphicx}
\usepackage{amssymb}
\usepackage{theorem}
\usepackage{float}
\usepackage{verbatim}

\floatstyle{ruled}
\newfloat{algorithm}{tbp}{loa}
\floatname{algorithm}{Algorithme}

\newtheorem{theo}{Th\'eor\`eme}
\newtheorem{defi}{D\'efinition}
\newenvironment{theorem}[1]{\vspace{-0.35cm}\begin{theo}#1}{\end{theo}\vspace{-0.3cm}}
\newenvironment{definition}[1]{\vspace{-0.35cm}\begin{defi}#1}{\end{defi}\vspace{-0.3cm}}

\author{Swan Dubois\addressmark{1}, Toshimitsu Masuzawa\addressmark{2} et S\'ebastien Tixeuil\addressmark{1}}

\title{Construction auto-stabilisante d'arbre couvrant en d\'epit d'actions malicieuses}

\address{
\addressmark{1}Universit\'e Pierre et Marie Curie \& INRIA (France),
 \{swan.dubois,sebastien.tixeuil\}@lip6.fr\\
\addressmark{2}Universit\'e d'Osaka (Japon),
masuzawa@ist.osaka-u.ac.jp
}


\begin{document}
\maketitle

\begin{abstract} 
Un protocole \emph{auto-stabilisant} est par nature tol\'erant aux fautes \emph{transitoires} (\emph{i.e.} de dur\'ee finie). Ces derni\`eres ann\'ees ont vu appara\^itre une nouvelle classe de protocoles qui, en plus d'\^etre auto-stabilisants, tol\`erent un nombre limit\'e de fautes \emph{permanentes}. Dans cet article, nous nous int\'eressons aux protocoles auto-stabilisants tol\'erant des fautes permanentes tr\`es s\'ev\`eres : les fautes \emph{byzantines}. Nous montrons que, pour certains probl\`emes n'admettant pas de solution \emph{strictement stabilisante} (\cite{NA02}), il existe toutefois des solutions tol\'erant des fautes byzantines si nous d\'efinissons un crit\`ere de tol\'erance moins contraignant.
\end{abstract}

\vspace{-0.2cm}
\section{Introduction}\label{sec:intro}

Le d\'eveloppement des syst\`emes distribu\'es \`a grande \'echelle a d\'emontr\'e que la tol\'erance aux diff\'erents types de fautes doit \^etre incluse dans les premi\`eres \'etapes du d\'eveloppement d'un tel syst\`eme. L'\emph{auto-stabilisation} permet de tol\'erer des fautes \emph{transitoires} tandis que la tol\'erance aux fautes traditionnelle permet de masquer l'effet de fautes \emph{permanentes}. Il est alors naturel de s'int\'eresser \`a des syst\`emes qui regrouperaient ces deux formes de tol\'erance. Cet article s'inscrit dans cette voie de recherche.

\paragraph{Auto-stabilisation} Dans cet article, nous consid\'erons un syst\`eme distribu\'e asynchrone anonyme, \emph{i.e.} un graphe non orient\'e connexe $G$ o\`u les sommets repr\'esentent les processus et les ar\^etes repr\'esentent les liens de communication. Deux processus $u$ et $v$ sont \emph{voisins} si l'ar\^ete $(u,v)$ existe dans $G$. Les variables d'un processus d\'efinissent son \emph{\'etat}. L'ensemble des \'etats des processus du syst\`eme \`a un instant donn\'e forme la \emph{configuration} du syst\`eme. Nous souhaitons r\'esoudre une classe particuli\`ere de probl\`emes sur ce syst\`eme : les probl\`emes \emph{statiques} (\emph{i.e.} les problèmes o\`u le syst\`eme doit atteindre un \'etat donn\'e et y rester). Par exemple, la construction d'arbre couvrant est un probl\`eme statique. De plus, nous considérons des probl\`emes pouvant \^etre sp\'ecifi\'es de mani\`ere locale (\emph{i.e.} il existe, pour chaque processus $v$, un pr\'edicat $spec(v)$ qui est vrai lorsque la configuration locale de $v$ est conforme au probl\`eme). Les variables apparaissant dans $spec(v)$ sont appel\'ees \emph{variables de sortie} ou  \emph{S-variables}.

Un syst\`eme auto-stabilisant (\cite{D74}) est un syst\`eme atteignant en un temps fini une configuration l\'egitime (\emph{i.e.} $spec(v)$ est vraie pour tout $v$) indépendament de la configuration initiale. Une fois cette configuration légitime atteinte, tout processus $v$ v\'erifie $spec(v)$ pour le restant de l'ex\'ecution (et donc, dans le cas d'un probl\`eme statique, le syst\`eme ne modifie plus ses S-variables). Par d\'efinition, un tel syst\`eme peut tol\'erer un nombre arbitraire de fautes \emph{transitoires}, \emph{i.e.} des fautes de dur\'ee finie (la configuration initiale arbitraire mod\'elisant le r\'{e}sultat de ces fautes). Cependant, la stabilisation du syst\`eme n'est en général garantie que si tous les processus ex\'ecutent correctement leur protocole. 

\paragraph{Stabilisation stricte} Si certains processus exhibent un comportement byzantin (\emph{i.e.} ont un comportement arbitraire, et donc potentiellement malicieux), ils peuvent perturber le syst\`eme au point que certains processus corrects ne v\'erifient jamais $spec(v)$. Pour g\'erer ce type de fautes, \cite{NA02} d\'efinit un protocole \emph{strictement stabilisant} comme un protocole auto-stabilisant tol\'erant des fautes byzantines permanentes. Pour en donner une d\'efinition formelle (dans le cas des probl\`emes statiques), nous devons introduire quelques notations et d\'{e}finitions.

Nous prenons comme mod\`ele de calcul le \emph{mod\`ele \`a \'etats} : Les variables des processus sont partag\'ees : chaque processus a un acc\`es direct en lecture aux variables de ses voisins. En une \emph{\'etape} atomique, chaque processus peut lire son \'etat et ceux de ses voisins et modifier son propre \'etat. Un \emph{protocole} est constitu\'e d'un ensemble de r\`egles de la forme $<garde>\longrightarrow<action>$. La $garde$ est un pr\'edicat sur l'\'etat du processus et de ses voisins tandis que l'$action$ est une s\'equence d'instructions modifiant l'\'etat du processus. A chaque \'etape, chaque processus \'evalue ses gardes. Il est dit \emph{activable} si l'une d'elles est vraie. Il est alors autoris\'e \`a ex\'ecuter son $action$ correspondante. Les \emph{ex\'ecutions} du syst\`eme (s\'equences d'\'etapes) sont g\'er\'ees par un \emph{ordonnanceur} : \`a chaque \'etape, il s\'electionne au moins un processus activable pour que celui-ci ex\'ecute sa r\`egle. Cet ordonnanceur permet de mod\'eliser l'asynchronisme du syst\`eme. La seule hypoth\`ese que nous faisons sur l'ordonnancement est qu'il est \emph{faiblement \'equitable}, \emph{i.e.} qu'aucun processus ne peut rester infiniment longtemps activable sans \^etre choisi par l'ordonnanceur.

\begin{definition} Un processus correct est \emph{$c$-correct} s'il est situ\'e \`a au moins $c$ sauts du byzantin le plus proche.
\end{definition}

\begin{definition}\label{def:cfcontained} Une configuration $\rho$ est \emph{$(c,f)$-contenue} pour $spec$ si, \'etant donn\'e au plus $f$ byzantins, tout processus $c$-correct $v$ v\'erifie $spec(v)$ et ne modifie pas ses S-variables dans toute ex\'ecution issue de $\rho$.
\end{definition}

Le param\`etre $c$ de la d\'efinition \ref{def:cfcontained} fait r\'ef\'erence au \emph{rayon de confinement} d\'efini dans \cite{NA02}. Le param\`etre $f$ fait r\'ef\'erence au nombre de byzantins alors que \cite{NA02} traite d'un nombre non born\'e de byzantins.

\begin{definition} Un protocole est \emph{$(c,f)$-strictement stabilisant} pour $spec$ si, \'etant donn\'e au plus $f$ byzantins, toute ex\'ecution (issue d'une configuration arbitraire) contient une configuration $(c,f)$-contenue pour $spec$.
\end{definition}

Une limite importante du mod\`ele de tol\'erance de \cite{NA02} est la notion de sp\'ecification $r$-\emph{restrictive}. Intuitivement, il s'agit d'une sp\'ecification interdisant des combinaisons d'\'etats de deux processus distants de $r$ sauts l'un de l'autre. Un r\'esultat important de \cite{NA02} est le suivant : s'il existe une solution $(c,f)$-strictement stabilisante \`a un probl\`eme admettant une sp\'ecification $r$-restrictive alors $c\geq r$. Pour certains probl\`emes, dont la construction d'arbre couvrant, on peut montrer que $r$ ne peut pas \^etre born\'e. En cons\'equence, il n'existe pas de solution $(c,f)$-strictement stabilisante pour tout rayon de confinement $c$ pour le problème de la construction d'un arbre couvrant.


\paragraph{Stabilisation forte} Pour contourner de tels r\'esultats d'impossibilit\'e, nous d\'efinissons ici un mod\`ele de tol\'erance plus faible : la \emph{stabilisation forte} (\cite{MT06}). Intuitivement, nous affaiblissons les contraintes relatives au rayon de confinement. En effet, nous autorisons certains processus \`a l'ext\'erieur de ce rayon \`a ne pas respecter la sp\'ecification en raison des byzantins. Cependant, ces perturbations sont limit\'ees dans le temps : les processus ne peuvent \^etre perturb\'es par les byzantins qu'un nombre fini de fois et toujours pendant un temps limit\'e m\^eme si les byzantins agissent infiniment longtemps. En voici la d\'efinition formelle dans le cas des probl\`emes statiques.

\begin{definition} Une configuration est \emph{$c$-l\'egitime} pour $spec$ si tout processus $c$-correct $v$ v\'erifie $spec(v)$.
\end{definition}

\begin{definition} Une configuration est \emph{$c$-stable} si tout processus $c$-correct ne modifie pas ses S-variables tant que les byzantins n'effectuent aucune action.
\end{definition}

\begin{definition} Une portion d'ex\'ecution $e=\rho_0,\rho_1,\ldots,\rho_t$ ($t>1$) est une \emph{$c$-perturbation} si : (1) $e$ est finie, (2) $e$ contient au moins une action d'un processus $c$-correct modifiant une S-variable, (3) $\rho_0$ est $c$-l\'egitime pour $spec$ et $c$-stable, et (4) $\rho_t$ est la première configuration $c$-l\'egitime pour $spec$ et $c$-stable apr\`es $\rho_0$.
\end{definition}

\begin{definition} Une configuration $\rho_0$ est \emph{$(t,k,c,f)$-temporellement contenue} pour $spec$ si, \'etant donn\'e au plus $f$ byzantins : (1) $\rho_0$ est $c$-l\'egitime pour $spec$ et $c$-stable, (2) toute ex\'ecution issue de $\rho_0$ contient une configuration $c$-l\'egitime pour $spec$ apr\`es laquelle les S-variables de tout processus $c$-correct ne sont pas modifi\'ees (m\^eme si les byzantins ex\'ecutent une infinit\'e d'actions), (3) toute ex\'ecution issue de $\rho_0$ contient au plus $t$ $c$-perturbations, et (4) toute ex\'ecution issue de $\rho_0$ contient au plus $k$ modifications des S-variables de chaque processus $c$-correct.
\end{definition}

Remarquons qu'une configuration $(t,k,c,f)$-temporellement contenue est $(c,f)$-contenue si $t=k=0$. Le premier concept est donc une g\'en\'eralisation du second.

\begin{definition} Un protocole $\mathcal{P}$ est \emph{$(t,c,f)$-fortement stabilisant} pour $spec$ si, \'etant donn\'e au plus $f$ byzantins, toute ex\'ecution (issue d'une configuration arbitraire) contient une configuration $(t,k,c,f)$-temporellement contenue pour $spec$ atteinte en au plus $l$ unit\'es de temps. $l$ et $k$ d\'esignent respectivement le \emph{temps de stabilisation} et le \emph{temps de perturbation} de $\mathcal{P}$.
\end{definition}
\newpage
Par d\'efinition, un protocole fortement stabilisant est plus faible qu'un protocole strictement stabilisant (car un processus en dehors du rayon de confinement peut subir l'influence des byzantins). Cependant, il est plus puissant qu'un protocole auto-stabilisant (qui peut ne jamais stabiliser en pr\'esence de byzantins).

\paragraph{Discussion} Il existe une analogie entre la puissance respective de la $(c,f)$-stabilisation stricte et de la $(t,k,c,f)$-stabilisation forte d'une part et l'auto-stabilisation et la pseudo-stabilisation d'autre part. Un protocole \emph{pseudo-stabilisant} (\cite{BGM93}) garantit que toute ex\'ecution (issue d'une configuration arbitraire) a un suffixe qui v\'erifie la sp\'ecification. Cependant, il est possible qu'une ex\'ecution n'atteigne jamais une configuration l\'egitime \`a partir de laquelle toute ex\'ecution v\'erifie la sp\'ecification. En effet, un protocole pseudo-stabilisant peut se comporter suivant la sp\'ecification mais en ayant la possibilité de l'invalider dans le futur. Un ordonnancement particulier peut emp\^echer un tel protocole d'avoir un comportement correct pendant un temps arbitrairement long. Toutefois, un protocole pseudo-stabilisant peut \^etre utile car un comportement correct est ultimement atteint. De mani\`ere similaire, toute ex\'ecution d'un protocole $(t,k,c,f)$-fortement stabilisant a un suffixe tel que tout processus $c$-correct a un comportement correct. Cependant, ce protocole peut ne jamais atteindre  une configuration apr\`es laquelle les byzantins ne peuvent plus perturber les processus $c$-corrects. En effet, tous les processus $c$-corrects peuvent avoir un comportement correct pendant un temps arbitrairement long tout en ayant la possibilit\'e d'effectuer au plus $k$ actions incorrectes (en au plus $t$ perturbations au niveau du syst\`eme). Une diff\'erence importante (mais subtile) est que les perturbations d'un protocole fortement stabilisant sont d\^ues uniquement aux byzantins alors que les invalidations de sp\'ecification d'un protocole pseudo-stabilisant sont d\^ues \`a l'ordonnancement.

\vspace{-0.2cm}
\section{Construction d'arbre couvrant}\label{sec:arbre}

\paragraph{Probl\`eme} Dans cette section, nous nous int\'eressons au probl\`eme de la construction d'un arbre couvrant du syst\`eme. Il s'agit d'un probl\`eme fondamental car il permet de mettre en \oe uvre de nombreux protocoles de communication (par exemple, diffusion, routage par les plus courts chemins, etc.).

Nous consid\'erons que le syst\`eme est \emph{semi-uniforme} (il existe un processus $r$ distingué comme la racine de l'arbre \`a construire). Chaque processus $v$ dispose de deux S-variables : $P_v$ et $H_v$ qui d\'esignent respectivement le parent et la hauteur de $v$ dans l'arbre en construction. Le but de la construction est que l'ensemble des pointeurs $P_v$ forme un arbre couvrant du syst\`eme enracin\'e en $r$.
Nous consid\'erons un syst\`eme dans lequel un certain nombre de processus peuvent \^etre byzantins et donc exhiber un comportement arbitraire (ces processus peuvent donc se comporter comme des n\oe uds internes de l'arbre ou encore comme le processus racine). C'est pourquoi nous devons faire certaines hypothèses sur le syst\`eme. La premi\`ere est que le processus $r$ est toujours correct (il n'aura jamais de comportement byzantin). La seconde est que l'ensemble des processus corrects reste toujours connect\'e. En d'autres termes, les byzantins ne partitionnent jamais le sous-syst\`eme des processus corrects.

Dans ces conditions, il est impossible, pour un processus correct, de distinguer la racine r\'eelle $r$ d'un byzantin se comportant comme une racine. Nous devons donc autoriser le syst\`eme \`a construire une for\^et couvrante du syst\`eme (donc un ensemble d'arbres couvrant le syst\`eme) dans laquelle chaque racine est soit $r$ soit un byzantin. Voici la sp\'ecification formelle du probl\`eme que nous consid\`erons dans cet article :
\vspace{-0.2cm}
\[spec(v) : \begin{cases}
 (P_v=\bot) \wedge (H_v = 0) \text{ si } v \text{ est la racine } r \\
 (P_v \in N_v) \wedge (P_v \text{ correct}\Rightarrow H_v = H_{P_v}+1) \text{ dans le cas contraire}
\end{cases}\]
\vspace{-0.2cm}

Il est possible de remarquer que, dans le cas o\`u aucun processus n'est byzantin et o\`u tout processus $v$ v\'erifie $spec(v)$, il existe un arbre couvrant du syst\`eme au sens ``classique''. De plus, il faut noter que cette sp\'ecification n'impose aucune contrainte sur l'arbre construit (arbre en largeur, degr\'e des n\oe uds...).

\begin{algorithm}\label{algo}
\caption{$\mathcal{CAFS}$: Construction d'arbre couvrant fortement stabilisante pour le processus $v$.}
\begin{tabbing}
xxx \= xxx \= xxx \= xxx \= xxx \= \kill
\small\textbf{Constantes :}\\
\> \small$\Delta_v$ le degr\a'e de $v$ \\
\> \small$N_v$ l'ensemble des voisins de $v$ (pour $k\in N_v$, le num\a'ero de canal de $k$ est not\a'e $\Vert k \Vert$) \\
\small\textbf{S-variables :}\\
\> \small$P_v\in N_v\cup\{\bot\}$ : parent de $v$\\
\> \small$H_v\in \mathbb{N}$ : hauteur de $v$ \\
\small\textbf{R\a`egles :}\\
\> \small$(v=r)\wedge((P_v \neq \bot)\vee(H_v \neq 0))\longrightarrow H_v :=0 ;~P_v := \bot$\\
\> \small$(v\neq r)\wedge((P_v \notin N_v)\vee(H_v\neq H_{P_v}+1))\longrightarrow P_v := suivant_v(P_v) ;~H_v := H_{P_v}+1 \text{ o\a`u } \Vert suivant_v(k) \Vert=(\Vert k \Vert +1)\text{ mod }\Delta_v$
\end{tabbing}
\end{algorithm}

\paragraph{Solution fortement stabilisante} Dans la majorit\'e des constructions d'arbre couvrant auto-stabilisantes, chaque processus v\'erifie localement la coh\'erence de sa hauteur par rapport \`a celle de ses voisins. Quand il d\'etecte une incoh\'erence, il modifie sa variable parent pour choisir un ``meilleur'' voisin. Le crit\`ere de qualit\'e d'un voisin d\'epend en r\'ealit\'e de la propri\'et\'e globale souhait\'ee pour l'arbre. Lorsque que le syst\`eme contient des processus byzantins, ceux-ci peuvent perturber un nombre non born\'e de processus en prenant successivement des états ``meilleurs'' et ``pires'' que leurs voisins. C'est pourquoi le protocole que nous proposons ici suit une autre approche. L'id\'ee principale est de contourner ce genre de perturbations par la strat\'egie suivante : lorsqu'un processus d\'etecte une incoh\'erence locale, il ne choisit pas un ``meilleur'' voisin mais choisit le voisin suivant son parent actuel selon un ordre cyclique sur ses voisins. 

L'algorithme 1 pr\'esente notre protocole de construction d'arbre couvrant fortement stabilisant $\mathcal{CAFS}$ qui peut tol\'erer un nombre arbitraire de byzantins (autres que la racine) tant que le sous ensemble des processus corrects reste connexe. Son rayon de confinement est \'egal \`a $0$ ce qui est \'evidement optimal.

Intuitivement, la correction de ce protocole repose sur les \'el\'ements suivants. Premi\`erement, tout processus correct $v$ est activable si et seulement si $spec(v)$ est faux. Cela signifie que toute configuration dans laquelle aucun processus correct n'est activable est $0$-l\'egitime et $0$-stable. Deuxi\`emement, il faut remarquer que toute ex\'ecution issue d'une configuration dans laquelle au moins un processus correct $v$ ne v\'erifie pas $spec(v)$ atteint en un temps fini une configuration dans laquelle tout processus correct $v$ v\'erifie $spec(v)$. Pour cela, il faut constater que la racine $r$ fait au plus une action dans toute ex\'ecution. Ensuite, chaque voisin correct de $r$ fait au plus $2\Delta$ actions dans toute ex\'ecution o\`u $\Delta$ est le degr\'e du syst\`eme (\emph{i.e.} le degr\'e maximal de ses processus). Enfin, nous pouvons g\'en\'eraliser ce raisonement et dire que tout processus correct $v$ effectue au plus $O(\Delta^\delta)$ actions dans toute ex\'ecution o\`u $\delta$ est la distance entre $v$ et $r$ dans le sous-syst\`eme des processus corrects. Ceci permet de d\'eduire que le syst\`eme atteindra toujours en un temps fini une configuration dans laquelle tout processus correct $v$ v\'erifie $spec(v)$ (que ce soit en partant d'une configuration initiale ou apr\`es une action byzantine). Cela permet de conclure que le syst\`eme ne peut conna\^itre qu'un nombre fini de $0$-pertubations, chacune ayant une dur\'ee finie. Le th\'eor\`eme 1 r\'esume les propri\'et\'es de $\mathcal{CAFS}$.


\begin{theorem}Notons $d$ le diam\`etre du sous-syst\`eme constitu\'e des processus corrects et $f$ le nombre de byzantins. Le protocole $\mathcal{CAFS}$ est un protocole $(n\Delta^d, 0, n-1)$-fortement stabilisant pour la construction d'arbre couvrant. Le temps de stabilisation de $\mathcal{CAFS}$ est en $O((n-f)\Delta^d)$ \'etapes (de processus corrects) et le temps de perturbation de $\mathcal{CAFS}$ est $\Delta^d$.
\end{theorem}

\vspace{-0.2cm}
\section{Conclusion}\label{sec:conclusion}

Nous avons \'etudi\'e la classe des protocoles auto-stabilisants tol\'erant de plus des fautes byzantines permanentes. Cette classe contient les protocoles strictement stabilisants et fortement stabilisants. Par l'\'etude du probl\`eme de la construction d'arbre couvrant, nous avons illustr\'e le fait que les seconds permettent de r\'esoudre un plus grand nombre de probl\`emes que les premiers. En contrepartie, les propri\'et\'es de tol\'erance atteintes sont plus faibles. Une voie de recherche future possible est la caract\'erisation de la classe des probl\`emes admettant une solution fortement stabilisante mais pas de solution strictement stabilisante.

\vspace{-0.2cm}
\bibliographystyle{alpha}
\small{
\bibliography{Biblio}
}

\end{document}